\documentclass[twocolumn,english,prl,longbibliography,superscriptaddress]{revtex4-1}
\usepackage[colorlinks=true,urlcolor=blue,citecolor=blue,linkcolor=blue]{hyperref} 
\usepackage[T1]{fontenc}
\usepackage[latin9]{inputenc}
\usepackage{amssymb}
\usepackage{graphicx}
\usepackage{amsmath,color}
\usepackage{mathrsfs}
\usepackage{float}
\usepackage{indentfirst}
\usepackage{txfonts}

\makeatletter

\def\journal #1, #2, #3, 1#4#5#6{{\sl #1~}{\bf #2}, #3 (1#4#5#6) }

\def\eqa{\begin{eqnarray}}
\def\eea{\end{eqnarray}}
\newcommand{\eq}{\begin{equation}}
\newcommand{\ee}{\end{equation}}

\newcommand{\Eq}[1]{Eq.~(\ref{#1})}

\newcommand{\Tr}{{\rm Tr}}

\newcommand{\betaVQE}{$\beta$-VQE }

\makeatother

\usepackage{babel}

\begin{document}

\title{Solving Quantum Statistical Mechanics with \\ Variational Autoregressive Networks and Quantum Circuits}

\author{Jin-Guo Liu}
\affiliation{Institute of Physics, Chinese Academy of Sciences, Beijing 100190, China}

\author{Liang Mao}
\affiliation{Department of Physics, Tsinghua University, Beijing 100084, China}

\author{Pan Zhang}
\affiliation{Institute of Theoretical Physics, Chinese Academy of Sciences, Beijing 100190, China}

\author{Lei Wang}
\affiliation{Institute of Physics, Chinese Academy of Sciences, Beijing 100190, China}
\affiliation{Songshan Lake Materials Laboratory, Dongguan, Guangdong 523808, China}

\begin{abstract} 
We extend the ability of unitary quantum circuits by interfacing it with classical autoregressive neural networks. The combined model parametrizes a variational density matrix as a classical mixture of quantum pure states, where the autoregressive network generates bitstring samples as input states to the quantum circuit. We devise an efficient variational algorithm to jointly optimize the classical neural network and the quantum circuit for quantum statistical mechanics problems. One can obtain thermal observables such as the variational free energy, entropy, and specific heat. As a by product, the algorithm also gives access to low energy excitation states. We demonstrate applications to thermal properties and excitation spectra of the quantum Ising model with resources that are feasible on near-term quantum computers. 
\end{abstract}
\maketitle

\paragraph{Introduction--}
Quantum statistical mechanics poses two sets of challenges to classical computational approaches. First of all, classical algorithms generally encounter the difficulties of diagonalzing exponentially large Hamiltonians or the sign problem originates from the quantum nature of the problem. Moreover, even on the eigenbasis one still faces intractable partition function which involves summation of exponentially large number of terms.

A straightforward way to address these difficulties is to directly realize the physical Hamiltonian on analog quantum devices and study the system at thermal equilibrium, for example, see Refs.~\cite{VanHoucke2012, King2018}. 
On the other hand, a potentially more general approach would be to study thermal properties with a universal gate model quantum computer. 
However, it calls for algorithmic innovations to prepare thermal quantum states on quantum circuits given their unitary nature. There have been quantum algorithms to prepare thermal Gibbs states on quantum computers~\cite{Terhal2000, Poulin2009, Temme2011, Riera2012, Brandao2019}. Unfortunately, these approaches may not be feasible on near-term noisy quantum computers with limited circuit depth. While variational quantum algorithm for preparing thermofield double states~\cite{Wu2018k, Zhu2019} requires additional quantum resources such as ancilla qubits, as well as measuring  and extrapolating Renyi entropies. The quantum imaginary-time evolution~\cite{Motta2019} relies on exponentially difficult tomography on a growing number of qubits and synthesize of general multi-qubit unitaries.

Recently, Refs.~\cite{Martyn2019, Verdon2019a} proposed practical approaches to prepare the thermal density matrix as a classical mixture of quantum pure states in the eigenbasis. In these proposals, the classical probabilistic model is either assumed to be factorized or expressed as an energy-based model~\cite{Goodfellow-et-al-2016}. However, the factorized distribution is generally a crude approximation for the Gibbs distribution in the eigenbasis. 
While the energy-based model still faces the problem of intractable partition function, which inhibits efficient and unbiased sampling, learning, or even evaluating the model likelihood. 

\begin{figure}[t]
\includegraphics[width=\columnwidth]{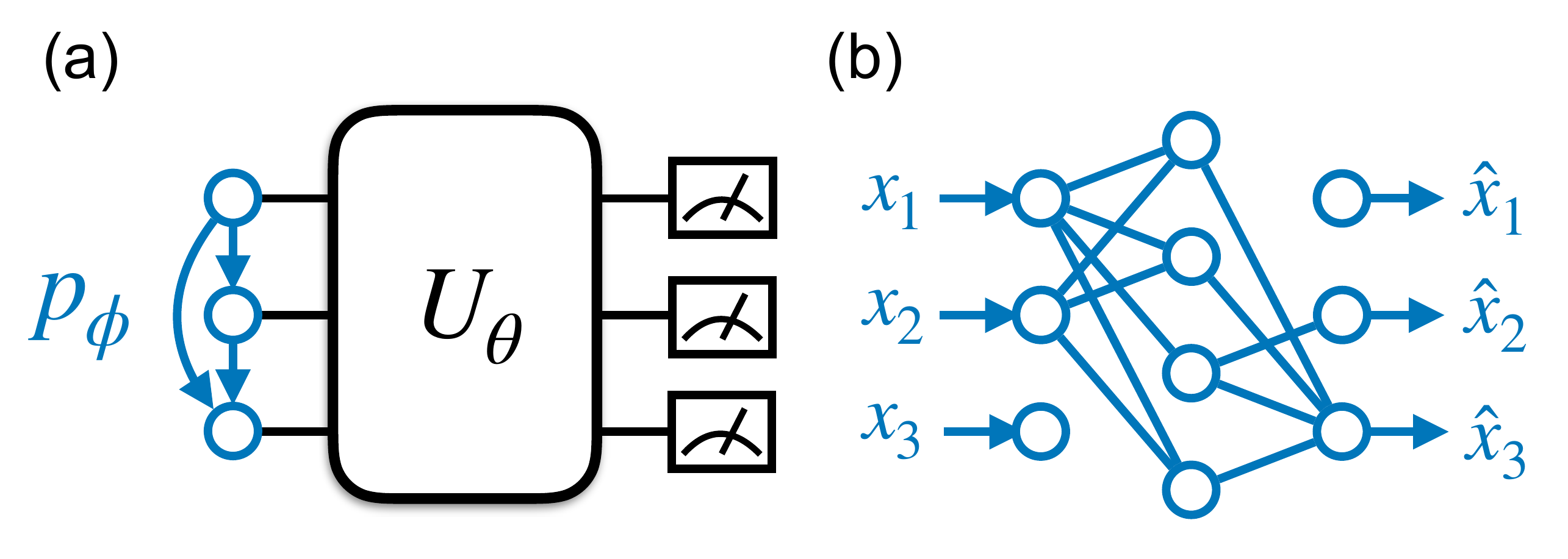}
\caption{(a) The autoregressive network shown in blue is a classical probabilistic model that parametrizes a joint distribution in the form of \Eq{eq:van}. The model generates bit string as easy to prepare input product states to the quantum circuit. The neural network and the circuit produce a parametrized density matrix \Eq{eq:dm}. (b) An implementation of the autoregressive model $p_\phi$ using the masked autoencoder~\cite{Germain2015}. The neural network maps bit strings to real-valued outputs which parametrizes the conditional probabilities in \Eq{eq:van}.}
\label{fig:setup}
\end{figure}

Modern probabilistic generative models offer solutions to the intractable partition function problem~\cite{PILtutorial} since the goals of generative modeling are exactly to represent, learn and sample from complex high-dimensional probability distributions efficiently. Popular generative models include autoregressive models~\cite{Germain2015, Oord2016, Oord2016b}, variational autoencoders~\cite{Kingma2013}, generative adversarial networks~\cite{Goodfellow2014a}, and flow-based models~\cite{Kobyzev2019}. For the purpose of this study, the autoregressive models stand out since they support unbiased gradient estimator for discrete variables, direct sampling, and tractable likelihood \emph{at the same time}. The autoregressive models have reached state-of-the-art performance in modeling realistic data and found real-world applications in synthesizing natural speech and images~\cite{Oord2016, Oord2016b}. Variational optimization of the autoregressive network has been used for classical statistical physics problems~\cite{Wu2018f, Pan2019}. Quantum generalization of the network was also employed for ground state of quantum many-body systems~\cite{Sharir2019}. 

In this paper, we combine quantum circuits with autoregressive probabilistic models to solve problems in quantum statistical mechanics. The resulting model allows one to perform variational free energy over density matrices efficiently. We demonstrate applications of the approach to thermal properties and excitations of quantum lattice model. 

By leveraging the recent advances in deep probabilistic generative models, the proposed approach extends the variational quantum eigensolver (VQE)~\cite{Peruzzo2014} to thermal quantum states with essentially no overhead. Thus, the present algorithm is also feasible for near-term quantum computers~\cite{Shen2017b, OMalley2016, Kandala2017, Colless2018, Hempel2018, Kokail2019, Mazzola2019}. The only practical difference to the VQE is that one needs to sample input states to the quantum circuit from a classical distribution, and one has an additional term in the objective function to account for the entropy of the input distribution. 

For the classical simulation of the proposed algorithm, we use \texttt{Yao.jl}, an extensible and efficient framework for quantum algorithm design~\cite{Yao.jl}. \texttt{Yao.jl}'s batched quantum register and automatic differentiation via reversible computing~\cite{YaoPaper} makes it an ideal tool for differentiable programming models which combine classical neural networks and quantum circuits. Our code implementation can be found at ~\cite{github}. 


\paragraph{Model architecture, objective function, and optimization scheme--} Figure~\ref{fig:setup}(a) shows the architecture of the variational ansatz. A classical probabilistic model generates binary random variables $x$ according to a classical distribution $p_\phi(x)$, where $\phi$ are the network parameters. It is straightforward to prepare qubits to the classical product state $|x\rangle$. Then, a parametrized quantum circuit performs unitary transformation to the input states $U_\theta | x\rangle$, where the circuit parameters $\theta$ do not depend on the inputs. Overall, the model produces a classical mixture of quantum states. The density matrix of the ensemble reads~\cite{Martyn2019, Verdon2019a}
\begin{equation}
\rho=\sum_{x} p_{\phi}(x) U_{\theta}|x \rangle  \langle x| U_{\theta}^{\dagger}. 
\label{eq:dm}
\end{equation}
The density matrix is hermitian and positive definite. Moreover, given a normalized classical probability, one has $\Tr(\rho) = \sum_x p_\phi(x) = 1$. The density matrix depends both on parameters $\phi$ and $\theta$. We omit the explicit dependence in the notation to avoid cluttering in the notations. 

The parametrized quantum circuit performs a unitary transformation to the diagonal density matrix $\sum_x p_\phi (x) |x\rangle \langle x |$, whose diagonal elements are parametrized by a neural network. Using a quantum circuit for the unitary transformation~\cite{Verstraete2009} is more general than the classical flow model~\cite{Cranmer2019}. Moreover, it automatically ensures physical constraints such as the orthogonality of the eigenstates. The classical distribution $p_\phi(x)$ is in general nontrivial since it is not necessarily factorized for each dimension of $x$~\cite{Martyn2019, Verdon2019a}. Thus, exact representation of the classical distribution on the eigenbasis $p_\phi(x)$ may also incur exponential resources. Parametrizing the probability distribution using a classical Boltzmann distribution has the problem of intractable partition functions. Hence, we employ an autoregressive network to produce the input states of the quantum circuit. 

The autoregressive network models the joint probability distribution as a product of conditional probabilities 
\begin{align}
p_\phi(x) = \prod_i p_\phi(x_i | x_{<i}) = p_\phi(x_1) p_\phi(x_2| x_1) p_\phi(x_3| x_1, x_2) \ldots,
\label{eq:van}
\end{align}
where one has assumed an order of each dimension of the variables. $x_{<i}$ denotes the set of variables that are before $x_i$. 
The autoregressive network is a special form of Bayesian network, which models conditional dependence of random variables as a directed acyclic graph shown in Fig.~\ref{fig:setup}(a). The model can capture high-dimensional multimode distribution with complex correlations. One can also directly draw uncorrelated samples from the joint distribution via ancestral sampling, which follows the order of the conditional probabilities. 

The practical implementation of the autoregressive networks largely benefits from rapid development of deep learning architectures such as the recurrent or convolutional neural networks~\cite{Oord2016,Oord2016b} and autoencoders~\cite{Germain2015}.  In this paper, we employ the masked autoencoder shown in Fig.~\ref{fig:setup}(b). The autoencoder network transforms bit string $x$ to real-valued vector $\hat{x}$ of the same dimension, where each element satisfies $0< \hat{x}_i <1$, e.g., outputs of sigmoid activation functions~\cite{Goodfellow-et-al-2016}. We mask out some connections in the autoencoder network so the connectivity ensures that $\hat{x}_i$ only depends on the binary variable $x_{<i}$. Thus, each element of the output defines a conditional Bernoulli distribution $p_{\phi}(x_i |x_{<i}) =  \hat{x}_i^{x_i}  (1-\hat{x}_i)^{1-x_i}$ for the binary variable $x_i$. In this way, the joint probability for all binary variables satisfies the autoregressive property \Eq{eq:van}. 
Since each conditional probability is normalized $\sum_{x_i} p_{\phi}(x_i| x_{<i})=1$, the joint distribution is normalized by construction. The probability distribution is parameterized by the network parameters $\phi$. 
In a simple limit where the network is disconnected,
$\hat{x} = \mathrm{sigmoid}(\phi)$ and one restores the product state ansatz considered in Refs.~\cite{Martyn2019, Verdon2019a}. 

Given a Hamiltonian $H$ at inverse temperature $\beta$, the density matrix $\sigma = e^{-\beta H}/Z$ plays a central role in the quantum statistical mechanics problem, where $Z=\operatorname{Tr}(e^{-\beta H})$ is an  intractable partition function. One can perform the variational calculation over the parametrized density matrix \Eq{eq:dm}  by minimizing the objective function
\begin{equation} 
\mathcal{L}= \operatorname{Tr}(\rho \ln \rho)+ \beta\operatorname{Tr}(\rho H), 
\label{eq:loss}
\end{equation}
which follows the Gibbs-Delbr\"uck-Moli\`ere variational principle of quantum statistical mechanics~\cite{huber1968variational}. The two terms of \Eq{eq:loss} correspond to the entropy and the expected energy  of the variational density matrix respectively. The objective function is related to the quantum relative entropy $S(\rho \| \sigma) = \mathcal{L} + \ln Z$ between the variational and the target density matrices. Since the relative entropy is nonnegative~\cite{Nielsen-Chuang}, one has $\mathcal{L} \ge - \ln Z$, i.e. the loss function is lower bounded by the physical free energy. The equality is reached only when the variational density matrix reaches the physical one $\rho =\sigma$.

To estimate the objective function \Eq{eq:loss}, one can sample a batch of input states $|x\rangle$ from the autoregressive network, then apply the parametrized circuit and measure the following estimator
\begin{equation}
\mathcal{L} = \mathbb{E}_{x\sim p_\phi(x)} \left[ \ln p_{\phi}(x)+ \beta \left\langle x\left|U_{\theta}^{\dagger} H U_{\theta}\right| x \right\rangle \right]. 
\label{eq:exp-loss}
\end{equation}
The first term depends solely on the classical probabilistic model, which can be directly computed via \Eq{eq:van} on the samples. Note that the entropy of the autoregressive model is known exactly rather than being intractable in the energy-based models~\cite{Verdon2019a}. Moreover, having direct access to the entropy avoids the difficulties of extrapolating the Renyi entropies measured on the quantum processor~\cite{Wu2018k, Zhu2019}. The second term of \Eq{eq:exp-loss} involves the expected energy of Hamiltonian operators $\langle H\rangle$, where we denote $\langle O \rangle = \mathbb{E}_{x\sim p_\phi(x)} \left[ \left\langle x\left|U_{\theta}^{\dagger} O U_{\theta}\right| x \right\rangle \right]$. The classical neural network and the quantum circuit perform classical and quantum average respectively. The \Eq{eq:exp-loss} shows zeros variance property, i.e. when the variational density matrix exactly reaches to the physical one, the variance of the estimator \Eq{eq:exp-loss} reduces to zero. This can be used as a self-verification of the variational ansatz and minimization procedure~\cite{Kokail2019}. 

We would like to utilize the gradient information to train the hybrid model which consists of neural networks and quantum circuits efficiently. Moreover, random sampling of the autoregressive net and the quantum circuit suggest that one should employ stochastic optimization with noisy gradient estimators~\cite{Goodfellow-et-al-2016}. First, the gradient with respect to the circuit parameters reads
\begin{equation}
\nabla_\theta \mathcal{L} =  \beta\, \mathbb{E}_{x \sim p_{\phi}} \left[ \nabla_\theta \left \langle x\left|U_{\theta}^{\dagger} H U_{\theta}\right|x \right \rangle  \right] \label{eq:grad_phi}. 
\end{equation}
The term  inside the square bracket is a gradient of a quantum expected value. To evaluate the expectation on an actual quantum device, one can employ the parameter shift rule of~\cite{Li2017b, Mitarai2018, Schuld2019, Nakanishi2019}. These approaches estimate the gradient of each circuit parameter using the difference of two sets of measurement on the quantum circuit with the same architecture. While in the classical simulation of the quantum algorithm one can employ the automatic differentiation~\cite{Baydin2018} to evaluate the gradient efficiently. 

\begin{figure}[t]
\includegraphics[width=\columnwidth,trim={0 1cm 0 0cm},clip]{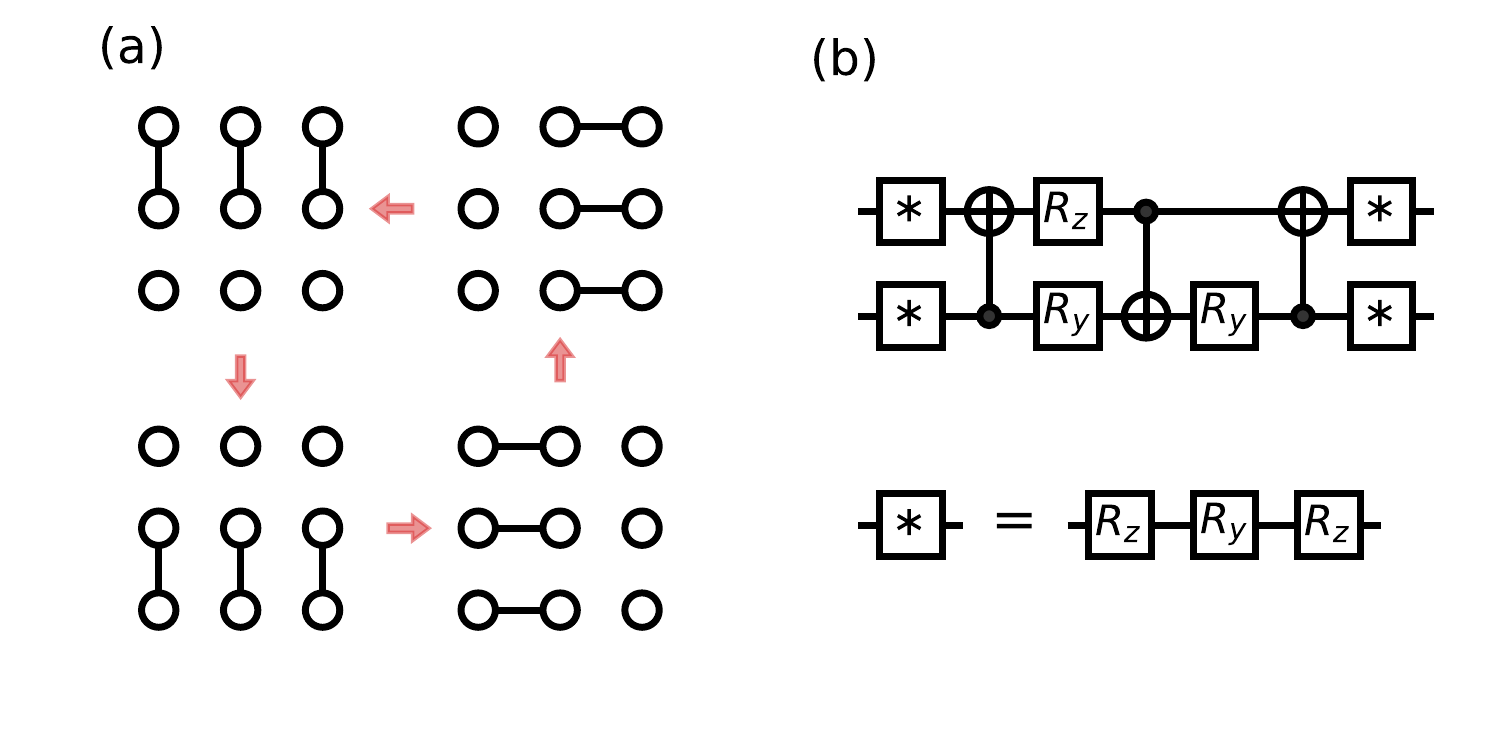}
\caption{(a) Layout of the two qubit unitaries in the variational circuit $U_\theta$. In each step one applies two-qubit  gates to adjacent qubits. (b) The two-qubit SU(4) gate consists of 3 CNOT and 15 single qubit rotation gates. Each single-qubit gate contains a learnable parameter~\cite{Shende2004}.}
\label{fig:circuit}
\end{figure}

Next, the gradients of the neural network parameters can be evaluated using the REINFORCE algorithm~\cite{Willia1992}
\begin{equation}
\nabla_\phi \mathcal{L} =  \mathbb{E}_{x \sim p_{\phi}} \left[ \left(f(x) -b\right)  \nabla_\phi \ln p_\phi(x) \right], \label{eq:grad_theta}  
\end{equation}
where the term $\nabla_\phi \ln p_\phi(x)$, known as the score-function gradient in the machine learning literature~\cite{Mohamed2019}, can be efficiently evaluated via backpropagation through the probabilistic model \Eq{eq:van}~\cite{Baydin2018}. In this regard, $f(x) =  \ln p_\phi(x) + \beta \langle x |U_{\theta}^{\dagger} H U_{\theta} |x \rangle$ can be viewed as the "reward signal" given the policy $p_\phi(x)$ for generating bit string samples. We have introduced the baseline $b = \mathbb{E}_{x \sim p_{\phi}} \left[f(x)\right]$ which does not affect the expectation of \Eq{eq:grad_theta} since $\mathbb{E}_{x \sim p_{\phi}} \left[ \nabla_\phi \ln p_\phi(x) \right]=0$. However, the baseline helps to reduce the variance of the gradient estimator~\cite{Mnih2014a}. 

Given the gradient information we train the autoregressive network and the quantum circuit jointly with the stochastic gradient descend method. The training procedure finds out the circuit $U_\theta$ which approximately diagonalizes the density matrix and brings the negative log-likelihood $-\ln p_{\phi}(x)$ closer to the energy spectrum of the system. 
In principle, the same circui can diagonalize the density matrices at all temperatures if  $U_\theta$ fully diagonalize the Hamiltonian.  
However, in the practical variational calculation, this does not need to be the case to achieve good variational free energy since the temperature selects the relevant low-energy spectra which contributes mostly to the objective function. 

After training, one can sample a batch of input states $|x\rangle$ and treat them as approximations of the eigenstates of the system. Since the unitary circuit preserves orthogonality of the input states, the sampled quantum states span a low energy subspace of the Hamiltonian. For example, measuring the expected energy $\langle x\left|U_{\theta}^{\dagger} H U_{\theta}\right|x \rangle$ reveals the excitation energies of the system. In this respect, the objective function \Eq{eq:loss} is related to the weighted subspace-search VQE algorithm for the excited states~\cite{Nakanishi2018}. Different from the weighted subspace-search VQE, a single physical parameter inverse temperature $\beta$ controls the relative weights on the input states. Adaptive sampling of the autoregressive model provides the correct weights that spans the relevant low energy space. Due to its close connection to the original VQE algorithm~\cite{Peruzzo2014}, we denote the present approach as the \betaVQE algorithm. While suppose the Hamiltonian is diagonal in the computational basis, i.e., a classical Hamiltonian, one can leave out the quantum circuit and the approach falls back to the variational autoregressive network approach of Ref.~\cite{Wu2018f}. In the classical limit it is also obvious that the autoregressive ansatz is advantageous than a simple product ansatz. 

\paragraph{Numerical simulations--}
  
\begin{figure}[t]
\includegraphics[width=\columnwidth]{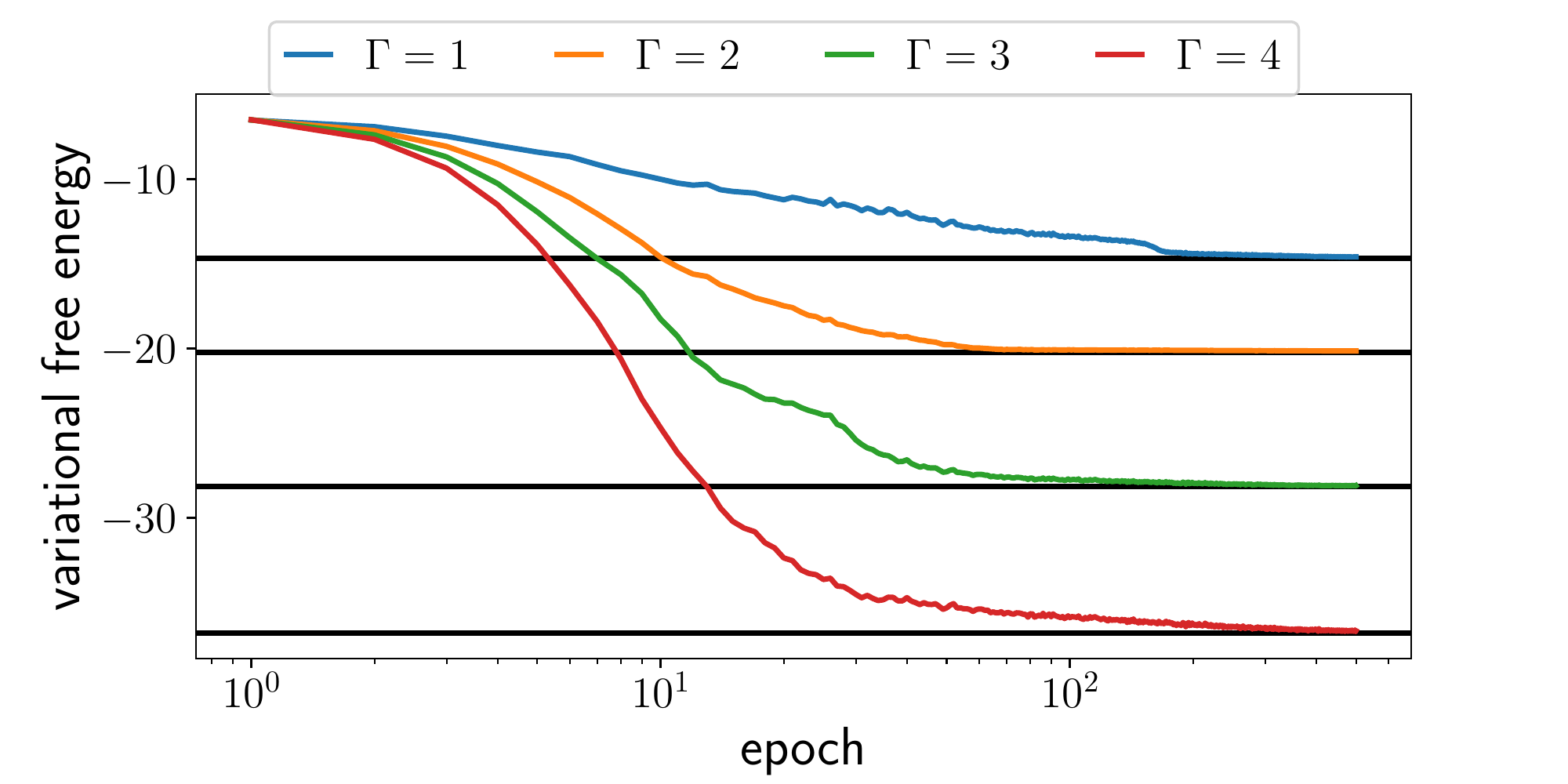}
\caption{The variational loss \Eq{eq:exp-loss} approaches to the exact free energies (black solid lines) of the $3\times 3$ quantum Ising model \Eq{eq:tfim} at $\beta=1$.} 
\label{fig:loss}
\end{figure}

We demonstrate application of \betaVQE to thermal properties of quantum lattice problems. 
Despite that most of the efforts on VQE have been devoted to quantum chemistry problems~\cite{Shen2017b, OMalley2016, Kandala2017, Colless2018, Hempel2018}, quantum lattice problems are more native applications on near-term quantum computers for two reasons. First, typical problems with local interaction one does not suffer from unfavorable scaling of a large number of Hamiltonian terms. Second, quantum lattice models that only involve spins and bosons do not invoke the overhead of mapping from fermion to qubits. Therefore, it is anticipated that near-term devices should already produce valuable results for quantum lattice problems before they are impactful for quantum chemistry problems~\cite{Wecker2015a}. 

We consider the prototypical transverse field  Ising model on a square lattice with open boundary conditions 
\begin{equation}
H = -\sum_{\langle i, j\rangle} Z_i Z_j - \Gamma \sum_i X_i,  
\label{eq:tfim}
\end{equation}
where $Z_i$ and $X_i$ are Pauli operators acting on the lattice sites. The model exhibits a quantum critical point at zero temperature at $\Gamma_c=3.04438(2)$. While for $\Gamma< \Gamma_c$ the model exhibits a thermal phase transition from an ferromagnetic phase to a disordered phase. All of these rich physics can be studied unbiasedly with sign-problem free quantum Monte Carlo approach, e.g. see~\cite{Hesselmann2016}. Having abundant established knowledge makes the problem \Eq{eq:tfim} an ideal benchmark problem for the \betaVQE algorithm on near-term quantum computers. 

\begin{figure}[t]
\includegraphics[width=\columnwidth]{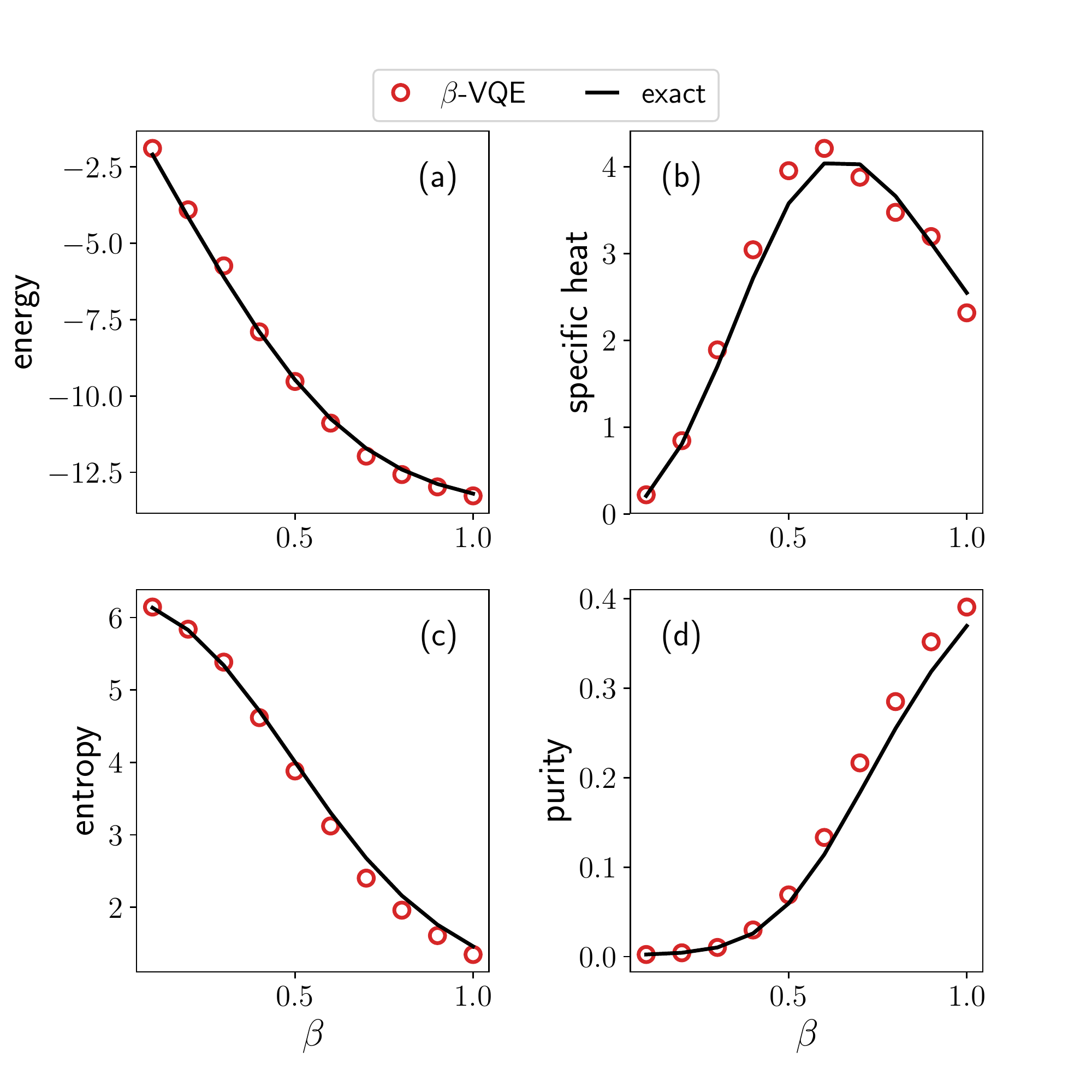}
\caption{Physical quantities obtained via the \betaVQE algorithm for the $3\times 3$ quantum Ising model \Eq{eq:tfim} at $\Gamma=3$. The black solid lines are exact solutions. 
}
\label{fig:obs}
\end{figure}

For the autoregressive network  \Eq{eq:van} we employ the masked autoencoder architecture~\cite{Germain2015} shown in Fig.~\ref{fig:setup}(b). We arrange the qubits on the two dimensional grid following the typewriter order. The autoencoder has single hidden layer of $500$ hidden neurons with rectified linear unit activation. 
For the variational quantum circuit, we employ the setup shown in Fig.~\ref{fig:circuit} which arrange the qubits on a two dimensional grid~\cite{Arute2019} and apply general two qubit gates~\cite{Shende2004} on the neighboring sites in each layer. The general gate consists of 15 single-qubit gates and 3 CNOT gates. Each two qubit unitary is parametrized by $15$ parameters in the rotational gates, which parametrizes the SU(4) group. The circuit architecture enjoys a balanced expressibility and hardware efficiency. We repeat the pattern for $d$ times which we denote as the depth $d$ of the variational quantum circuit. Therefore, for the $3\times 3$ system considered in Fig.~\ref{fig:circuit}(a) with $d=5$, there are $15 \times 12 \times  5=900$ circuit parameters. Initially we set all the circuit parameters to be zero. We estimate the gradients Eqs.~(\ref{eq:grad_phi}, \ref{eq:grad_theta}) on batch of 1000 samples, and we the Adam algorithms~\cite{Goodfellow-et-al-2016} to optimize the parameters $\phi$ and $\theta$ jointly. 

Figure~\ref{fig:loss} shows that the objective function decreases towards the exact values as a function of training epochs. We measure physical observables on the trained model and compare them with exact results. For example, Figs.~\ref{fig:obs}(a,b) show the expected energy $\langle H \rangle$ and the specific heat $ \beta^2\left( \langle H^2\rangle - \langle H\rangle^2\right)$ computed by measuring Hamiltonian expectation and its variance. Moreover, one sees in Fig.~\ref{fig:obs}(c) that the entropy $\mathbb{E}_{x\sim p_\phi(x)} \left[ -\ln p_{\phi}(x) \right]$ changes from $\ln 2$ per site in the high temperature limit towards zero at zero temperature. While the purity of the system $\operatorname{Tr}(\rho^2)=\mathbb{E}_{x\sim p_\phi(x)} \left[ p_{\phi}(x) \right]$ shown in  Fig.~\ref{fig:obs}(d) increases from zero towards one as the temperature decreases. All these observables can be directly measured on an actual quantum device. Overall, one sees the autoregressive model \Eq{eq:van} combined with variational quantum circuit yields accurate results over all temperatures. 

\begin{figure}[t]
\includegraphics[width=\columnwidth]{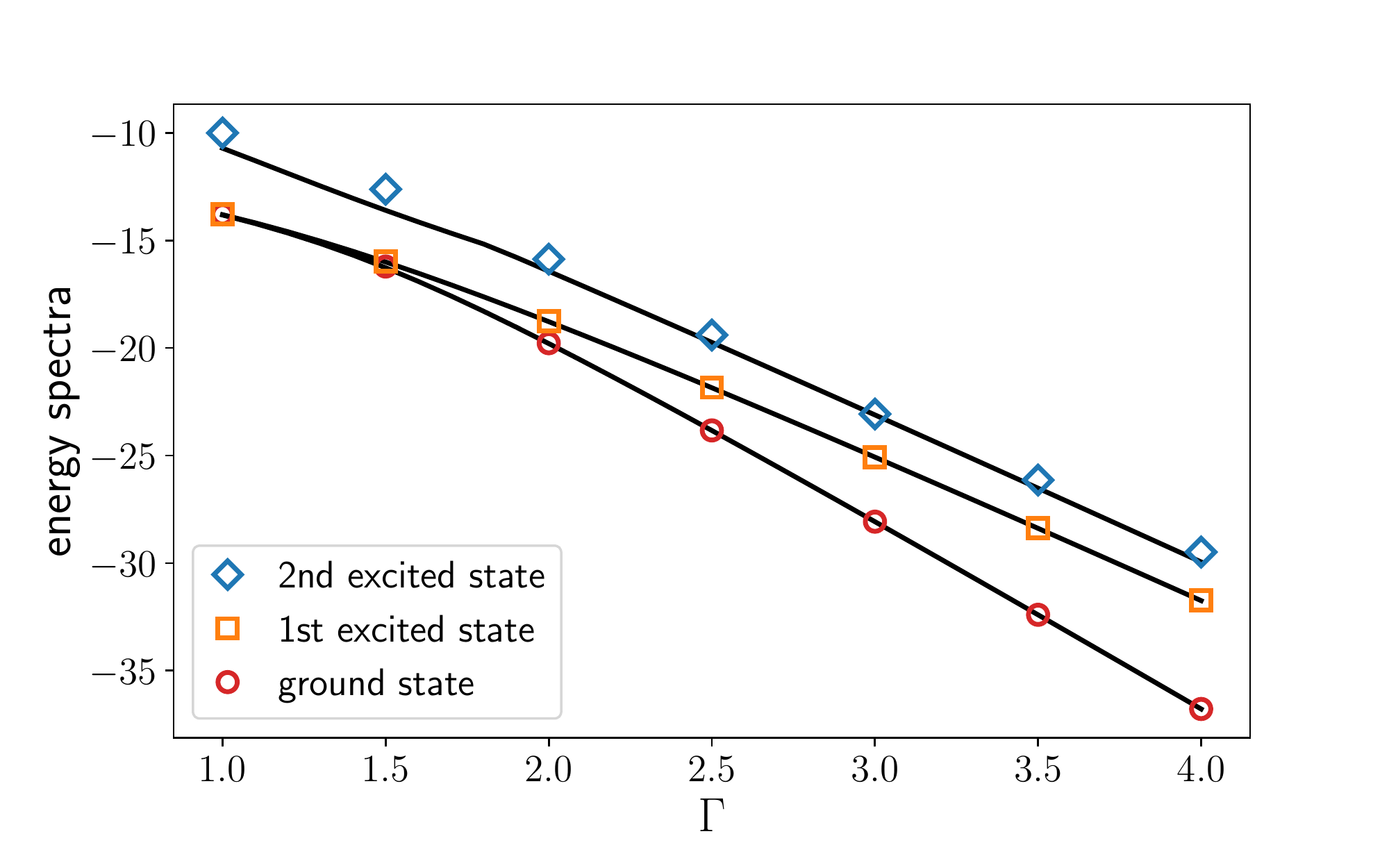}
\caption{Low energy excitation spectra of $3\times 3$ quantum Ising model. The symbols denote energy expectations $\left\langle x\left|U_{\theta}^{\dagger} H U_{\theta}\right| x \right\rangle$ measured on the samples of the autoregressive model trained at inverse temperature $\beta=0.5$. The solid lines are exact solutions.}
\label{fig:spectra}
\end{figure}


Figure~\ref{fig:spectra} shows the low energy spectrum of the quantum Ising model obtained from \betaVQE at $\beta =0.5$. One sees that the approach provides low energy spectrum of the problem at various strength of the transverse field. The approach works nicely even when the first excited state becomes nearly degenerated with the ground state. 

\paragraph{Outlooks--}
The present approach would be most useful for studying thermal properties of frustrated quantum systems which are prevented by the sign problem~\cite{Troyer2005}. Moreover, one can further employ the qubit efficient VQE scheme~\cite{Huggins2019, Liu2019c}, where one can study thermal properties of quantum many-body systems on a quantum computer with the number of qubits smaller than the number of degrees. In that scenario, the ansatz for the density matrix is a classical mixture of matrix product states. The variational ansatz for density matrix can also be used in quantum algorithm for non-equilibrium dynamics~\cite{Lamm2018a} and steady states~\cite{Yoshioka2019}.

The quantum circuit also acts as a canonical transformation that brings the density matrix to a diagonal representation. Combined with the fact that one can obtain the marginal likelihood of the leading bits in the autoregressive models, the setup may be useful for deriving effective models with less degrees of freedom similar to the classical case~\cite{Liw}. Therefore, one can envision using the present setup to derive effective models by using a quantum circuit for renormalization group transformation. Moreover, since the circuit approximately diagonalizes the density matrix, one also can make use of it for later purpose, such as accelerated time evolution~\cite{Cirstoiu2019}. 

Regarding further improvements of the algorithm, one may consider using tensor network probabilistic models~\cite{Han2018, Cheng2019b} instead of the autoregressive network to represent the classical distribution in the eigenbasis. Both models have the shortcoming that the sampling approach produces the bits sequentially. To address this issue, one may consider employ the recent proposed flow models for discrete variables~\cite{Tran2019a,Hoogeboom}. While to further improve the optimization efficiency, one may consider using the improved gradient estimator with even lower variances~\cite{Tucker2017b, Grathwohl2017}. To this end, differentiable programming of neural networks and quantum circuits shares a unified computational framework. Therefore, a seamlessly integration of models and techniques will enjoy advances of both worlds. 

\begin{acknowledgments}
\paragraph{Acknowledgment--}
We thank useful discussion with Nobuyuki Yoshioka and Yuya O. Nakagawa. We thank Xiu-Zhe Luo for contribution to \texttt{Yao.jl}~\cite{Yao.jl} and Dian Wu for collaboration in Ref.~\cite{Wu2018f}. The authors are supported by the National Natural Science Foundation of China under the Grant No.~11774398. 
\end{acknowledgments}

\bibliography{refs,manual_refs}

\end{document}